\documentclass[runningheads]{llncs}
\usepackage[utf8]{inputenc}

\usepackage{tabularx}
\usepackage{booktabs}
\usepackage{soul}
\usepackage{xcolor}
\usepackage{xspace}
\usepackage[hyphens]{url}
\usepackage{graphicx}
\usepackage[caption=false]{subfig}  
\usepackage[export]{adjustbox}
\usepackage{amsmath}
\usepackage{amssymb}
\usepackage{textcomp}
\usepackage{listings}
\usepackage{cleveref}
\usepackage{algorithm}
\usepackage[noend]{algpseudocode}
\usepackage{balance}
\usepackage{enumitem}
\usepackage{multirow}
\usepackage{colortbl}
\usepackage{balance}
\usepackage[flushleft]{threeparttable}

\definecolor{codegreen}{rgb}{0,0.6,0}
\definecolor{codegray}{rgb}{0.5,0.5,0.5}
\definecolor{codepurple}{rgb}{0.58,0,0.82}
\definecolor{backcolour}{rgb}{0.95,0.95,0.92}
\definecolor{gray}{gray}{0.9}

\lstdefinestyle{mystyle}{
    stringstyle=\color{deepgreen},
    backgroundcolor=\color{backcolour},   
    commentstyle=\color{codegreen},
    keywordstyle=\color{magenta},
    numberstyle=\tiny\color{codegray},
    stringstyle=\color{codepurple},
    basicstyle=\ttfamily\footnotesize,
    breakatwhitespace=false,         
    breaklines=true,                 
    captionpos=b,                    
    keepspaces=true,                 
    numbers=left,                    
    numbersep=5pt,                  
    showspaces=false,                
    showstringspaces=false,
    showtabs=false,                  
    tabsize=2
}

\newcommand{\HPLM}{High PLM\xspace}
\newcommand{\LPLM}{Low PLM\xspace}
\newcommand{\PL}{perceptual load\xspace}

\definecolor{APA_stats}{RGB}{100, 100, 120}
\newcommand{\APAstats}[2]{\textcolor{APA_stats}{(M = #1, SD = #2)}}

\newcommand{\APAt}[4]{\textcolor{APA_stats}{t(#1) = #2, p = #3; d = #4}}
\newcommand{\APAf}[5]{\textcolor{APA_stats}{F(#1, #2) = #3, p = #4, $\eta^{2}{p}$ = #5}}
\newcommand{\APAu}[3]{\textcolor{APA_stats}{U = #1, p = #2, CLES = #3}}

\newcounter{observation}
\newcommand{\observation}[1]{\refstepcounter{observation}
	\begin{center}
		\framebox{
			\begin{minipage}{0.9\columnwidth}
				{} \textit{#1}
			\end{minipage}
		}
	\end{center}
}

\providecommand{\keywords}[1]
{
  \small	
  \textbf{\textit{Keywords---}} #1
}

\begin{document}

\title{The Effect of Perceptual Load on Performance within IDE in People with ADHD Symptoms}

\titlerunning{Perceptual Load in IDE and Its Effect on People with ADHD Symptoms}
%

\author{Vseslav Kasatskii\inst{1} \and
Agnia Sergeyuk\inst{2} \and
Anastasiia Serova\inst{2} \and Sergey Titov\inst{2} \and Timofey Bryksin\inst{2}}
\authorrunning{V. Kasatskii et al.}
%
\institute{Neapolis University Pafos, Cyprus \and JetBrains Research\\
\email{kasatsky.vseslav@gmail.com,  \{agnia.sergeyuk, anastasiia.serova, sergey.titov, timofey.bryksin\}@jetbrains.com}}

\maketitle

\begin{abstract}
In this paper, we describe the research on how perceptual load can affect programming performance in people with symptoms of Attention Deficit / Hyperactivity Disorder (ADHD). We asked developers to complete the Barkley Deficits in Executive Functioning Scale, which indicates the presence and severity levels of ADHD symptoms. After that, participants solved mentally active programming tasks (coding) and monotonous ones (debugging) in the integrated development environment in high perceptual load modes (visually noisy) and low perceptual load modes (visually clear). The development environment was augmented with the plugin we wrote to track efficiency metrics, \textit{i.e.} time, speed, and activity. We found that the perceptual load does affect programmers' efficiency. For mentally active tasks, the time of inserting the first character was shorter and the overall speed was higher in the low perceptual load mode. For monotonous tasks, the total time for the solution was less for the low perceptual load mode. Also, we found that the effect of \PL on programmers' efficiency differs between those with and without ADHD symptoms. This effect has a specificity: depending on efficiency measures and ADHD symptoms, one or another level of perceptual load might be beneficial. Our findings support the idea of behavioral assessment of users for providing appropriate accommodation for the workforce with special needs.
\end{abstract}


\section{Introduction}\label{sec: introduction}

The diversity of employees in the rapidly changing and demanding field of software engineering is high. Partially, this diversity is dictated by the cognitive abilities and mental states of the workers. Due to the efforts of IT companies to be more inclusive, the neurodiversity of employees has become a focus of stakeholders and researchers in recent years~\cite{loiacono2018building,pisano2017neurodiversity,krzeminska,understanding,austin2017neurodiversity,dalton2013neurodiversity}.

Neurodiversity is the variability of mental functioning caused by neurodevelopmental disorders that cannot be cured but should be considered~\cite{dalton2013neurodiversity}. This means that people with such conditions can be socialized but need some accommodations at school and work. This paradigm was first applied to people with Autistic Spectrum Disorder~\cite{blume1998neurodiversity}, but now it also accounts for people with Attention-Deficit/Hyperactivity Disorder (ADHD), Dyslexia, etc..

The majority of existing research regarding accommodation for the differently abled workforce is focused on flexibility of work-related circumstances, social support, and coaching~\cite{telwatte2017workplace,mcgonagle2014coaching}. However, the field lacks thorough experimental works~\cite{doyle2020neurodiversity,reiss2004developmental}. To broaden the field of experimental research on work-station adjustments, we investigate whether and how integrated development environments (IDEs) can assist neurodiverse developers. 

There are studies that show that the task performance of neurodiverse people could be affected by the amount of information involved in the processing of the task~\cite{macdonald2011visual,carreiro2021adults,forster2014plugging}, \textit{i.e.} by the perceptual load of the environment and task. This is especially true for people with ADHD, who struggle with self-directed activities related to time management, Self-organization and problem-solving, Self-restraint, Self-motivation, and Self-regulation of emotions~\cite{barkley2011nature,barkley2013deficits,barkley2015history}. People with a higher frequency and severity of ADHD symptoms were found to have a better ability to inhibit unwanted processes, distract less, and therefore perform effectively when the perceptual load of the task was high~\cite{carreiro2021adults,forster2014plugging}. 

We wanted to test this effect in the field of programming, using an IDE as a medium. We conducted an experiment in which developers fill out the Barkley Deficits in Executive Functioning Scale (BDEFS-SF) to measure the existence and severity of ADHD symptoms in them. Then, they solved two types of programming tasks --- mentally active (coding) and monotonous (debugging). Each type of tasks was presented twice --- in the low and high perceptually loaded modes of IDE. In our research, the low perceptual load mode (\LPLM) has two interactable panels, while the high perceptual load mode (\HPLM) has seven. To manipulate the user interface and present tasks, we created a plugin, which also registered all indicators of solution efficiency.

We found that the mode of IDE and one's attentional difficulties affect the efficiency of programming. For mentally active tasks, the time of inserting the first character was significantly shorter and the solution speed was significantly higher in \LPLM compared to \HPLM (p = .005 for both metrics). For monotonous tasks, the total time for the solution and the time when the first bug was found were also significantly less in \LPLM than in \HPLM (p = .035 and p = .042, respectively). The effect of \PL differed across people with and without ADHD symptoms in our research. We found that the efficiency measure in question and the specificity of the symptoms might dictate the type of preferable accommodation. Namely, if the user has difficulties in self-organization, self-restraint, and self-regulation of emotions, and their effectiveness is measured as the activeness of the user, they may benefit from the \LPLM environment (p = .036, p = .022, p = .01, respectively). People with time-management impairment speed-wise benefit from \LPLM (p = .048). If one struggles with self-regulation of emotions and the measure of efficiency is time, one might benefit from \HPLM (p = .03). 

The results show that visual representation is an important aspect of IDE accessibility. A well-designed environment helps boost user productivity. Developing just-in-time work-station adjustments for neurodiverse people working in the programming field may support the efficiency of their performance.

\section{Background}\label{sec:back}

Research on the neurodiverse workforce in the field of software engineering highlights some challenges and counterbalancing advantages of such people. A literature review conducted by E. Costello and colleagues~\cite{costello2021professional} points out interpersonal and work-station usability challenges. From the interviews and survey conducted by Microsoft Research~\cite{understanding} it can be concluded that neurodiverse people who are involved in the field of programming usually struggle with focusing on tasks that are not interesting for them, with reviewing other peoples' code, with self-organization and self-motivation. Case studies, interviews, and blog posts of neurominorities show that in the field of software engineering, they are struggling with communication, self-organization, self-motivation, and focus on monotonous tasks~\cite{understanding,costello2021professional,reaktor}. This is usually dictated by the structure of the IT companies and the specificity of job tasks. Working with code in a noisy environment where there are social rules and various distractors may be not easy for anyone, but especially challenging for neurodiverse people. 

However, neurodiverse people also have strengths and unique traits that make them good candidates for positions in the field of software engineering. The advantages that a neurodiverse workforce inherits, such as the ability to notice patterns, creativity, hyperfocus, and adhering to rules, bring about a rise in team productivity and product quality along with other benefits for companies and employees~\cite{understanding,loiacono2018building}. These advantages might be less visible because of the difficulties people encounter. Therefore, to highlight their advantages and support their performance at work, neurodiverse people should be aided with different kinds of accommodations~\cite{motti2019designing,dong2013workplace}. 

According to the review by N. Doyle~\cite{doyle2020neurodiversity}, there are several categories of possible workplace adjustments for neurominorities: work environment flexibility; schedule flexibility; supervisor or co-worker support; support from different stakeholders; executive functions coaching; literacy coaching; training; assistive technologies and tools; work-station adjustments.

To provide appropriate tools to them, detailed research must be carried out; however, the field lacks thorough experimental works~\cite{doyle2020neurodiversity,reiss2004developmental}. The majority of existing research regarding accommodation for the differently abled workforce is focused on the flexibility of work-related circumstances, social support, and coaching~\cite{telwatte2017workplace,mcgonagle2014coaching}. In our work, our aim was to broaden the field of experimental research on work-station adjustments, investigating how an IDE might help neurodiverse developers cope with programming tasks.

In this paper, we focus on people with ADHD. ADHD is the diagnostic label for those who demonstrate significant problems with attention and / or impulsiveness and hyperactivity~\cite{antshel2006maternal}. According to the ADHD Institute~\cite{adhdinst}, the worldwide prevalence of ADHD in adults (18 to 44 years) has been estimated at 2.8\% overall (ranging from 0.6 to 7.3\% depending on the study)~\cite{fayyad2017descriptive}. In fact, the prevalence may be higher due to the underdiagnosis of this disorder~\cite{ginsberg2014underdiagnosis}. The percentage of adults who have the disorder can be shadowed by (a) lack of proper diagnostic tools, (b) lifetime adaptation to the symptoms. It is also known that people with neurodiversity are not always willing to disclose themselves at work~\cite{understanding,doyle2020neurodiversity,sumner2015neurodiversity}. For this reason, in the present work we are more oriented not on the diagnosis of ADHD itself in the sample, but on the specific results of the scale, measuring the severity of related symptoms. 

The scale we use in the present research is an empirically supported~\cite{barkley2010adhd}, valid~\cite{kamradt2021barkley} scale that was developed to assess the difficulties people with ADHD encounter. The scale in question is the Barkley Deficits in Executive Functioning Scale (BDEFS) and is based on the theory of ADHD~\cite{barkley2013deficits}, which was developed by Barkley~\cite{barkley1997behavioral}. This is the author whose work had an impact on the formation of diagnostic criteria for ADHD in the fifth edition of the Diagnostic and Statistical Manual of Mental Disorders (DSM-V)~\cite{barkley2007may,bell2011critical,american2013diagnostic}. 

The theory states that ADHD symptoms arise from biologically based difficulties that underpin the impairment of self-regulation of executive functions~\cite{barkley2015history}. In this context, executive functions are self-regulation abilities to internalize and use self-directed actions to choose goals and select, implement, and sustain actions toward those goals~\cite{antshel2014executive}. In other words, executive functions help people set goals and achieve them through self-regulation. Five self-regulatory self-directed activities accounted for in the Barkley's theory are~\cite{barkley2013deficits}: 
\begin{itemize}
    \item Self-inhibition --- the ability to stop one's reactions and behaviors, the capacity to prevent disruption of one's goal-oriented activity;
    \item Self-directed sensory-motor action --- polymodal nonverbal reexperiencing of past events, \textit{i.e.}, nonverbal working memory;
    \item Self-directed private speech --- internalized speech, verbal working memory;
    \item Self-directed emotion/motivation --- regulation of one's emotions and motivation using the first three executive functions;
    \item Self-directed play --- problem-solving with the help of analysis and synthesis.
\end{itemize}
Therefore, if one has persistent self-regulation difficulties that affect their adaptation to everyday life, we might suspect the disorder, which might be labeled as ADHD and that might imply the presence of neurodevelopmental differences. 

According to diagnostic criteria and based on the theory described above, people with ADHD symptoms might struggle the most with monotonous tasks~\cite{american2013diagnostic}. This is especially true for people with the inattention prevalence, which is known to persist in adults more than hyperactivity~\cite{american2013diagnostic}. Monotonous tasks generally require self-regulation, concentration, and maintaining attention, \textit{e.g.}, lengthy reading, thoroughly instructed activities. Software engineering might be considered such a task due to the repetitive, mentally demanding, and heavily regulated nature~\cite{hindle2016naturalness,couceiro2019pupillography}. This idea is also supported by the feedback of the workforce themselves~\cite{understanding,costello2021professional,reaktor}. This raises the question of how we can aid developers with ADHD symptoms in performing monotonous tasks during their work.

Previous research shows that one's ability to concentrate and self-regulate their attention can be affected by the perceptual load of the environment~\cite{lavie2005distracted}. The perceptual load is a limited amount of information that must be processed while performing a task~\cite{macdonald2011visual}. According to the integrated theory of structural and capacity approaches to attention~\cite{lavie1994perceptual}, if the perceptual load is high, distractors are not perceived, since most or all of the available capacity is exhausted. This might make it easier to concentrate on the task. 

It was noted in neuroimaging probes that patients with attention-related brain lesions showed a reduction in the capacity of attention and even a small increase in the perceptual load stopped them from being distracted~\cite{lavie2001role}. The same effect could be expected in neurodiverse people, since their brain is developmentally affected. Indeed, people with a higher frequency and severity of ADHD symptoms were found to have a better ability to inhibit unwanted processes and distract less when the perceptual load of the task was high~\cite{carreiro2021adults,forster2014plugging}.

To test whether the perceptual load of the IDE affects developers with ADHD symptoms in terms of the efficiency of solving programming tasks, we carried out an experiment, in which participants were asked to perform the most common programming tasks --- coding and debugging --- each in both low and high perceptually loaded IDE modes. With this experiment, our aim was to show if there is a possible way to compensate for attentional impairment with the help of software design tools. 

\section{Methodology}\label{sec:method}

Based on the theory and related work described in the previous section, we formulated the following research questions for programming tasks of different levels of attention activeness:

\textbf{RQ 1.} Does the \PL of the programming environment affect the efficiency of task solving?

\textbf{RQ 2.} Does the effect of \PL differ between people with and without impaired executive functions?

To find answers to these questions, we conducted experimental research, which is described in detail further in this section. 
All code for experiment replication and data analysis can be found in a public repository.\footnote{Data: \url{https://github.com/JetBrains-Research/adhd-study}}

\subsection{Sample}
The sample for our research consisted of adult developers --- students and practitioners --- who are familiar with the Python language. The participants were invited via e-mail and personal invitations across colleagues and acquaintances, using snowball sampling. We managed to gather 36 respondents.

\subsection{Materials}

\textbf{Barkley Deficits in Executive Functioning Scale.}

We used BDEFS-SF~\cite{ra2011barkley,sheble2019validation} as a valid self-assessment tool to measure the number and severity of difficulties with executive function in the sample. This multifactor scale allowed us to find specifically which executive functions might be deficient in participants.

BDEFS was first developed as a tool to perform a theory-based assessment of deficits in executive functioning in a clinical setting. Its early self-report version included five subscales labeled as five core dimensions of executive functions~\cite{barkley2011nature}: 
\begin{itemize}
    \item Self-Management to Time --- a sense of time, time-scheduling; 
    \item Self-Organization / Problem-Solving --- arrangement of one’s activities, coming up with ideas and solutions to encountered problems; 
    \item Self-Discipline (Restraint or Inhibition) --- the ability to inhibit one's reactions, consider consequences, self-awareness; 
    \item Self-Motivation --- the ability to cope with indolence and boredom, fulfilling duties without external motivators; 
    \item Self-Activation / Concentration --- sustaining attention and keeping up at work and in boring, uninteresting activities. 
\end{itemize}

This scale was tested in several studies in adults with ADHD and proved to be effective in distinguishing nonclinical and clinic-referred participants~\cite{barkley2010adhd}. Further work with the obtained data enabled Barkley and colleagues to notice and fix the underrepresentation of Self-regulation of the emotion domain in the scale. This resulted in the scale that tests one's executive functions related to Time management, Self-organization and problem-solving, Self-restraint, Self-motivation, and Self-regulation of emotions. This scale is now widely used in research such as ours, as well as in clinical settings. 

While filling out the scale, people were asked to indicate from 1 (never or rarely) to 4 (very often) how often they encounter one or another problem in their day-to-day life.

The scale enabled us to not fixate on labels and diagnoses, which represent sensitive data and gather those who were not necessarily diagnosed with ADHD but may have still attentional difficulties~\cite{ra2011barkley}.

\medskip
\textbf{Tasks.}

    \begin{figure}
    \centering
    \subfloat{\includegraphics[width = 0.57\textwidth]{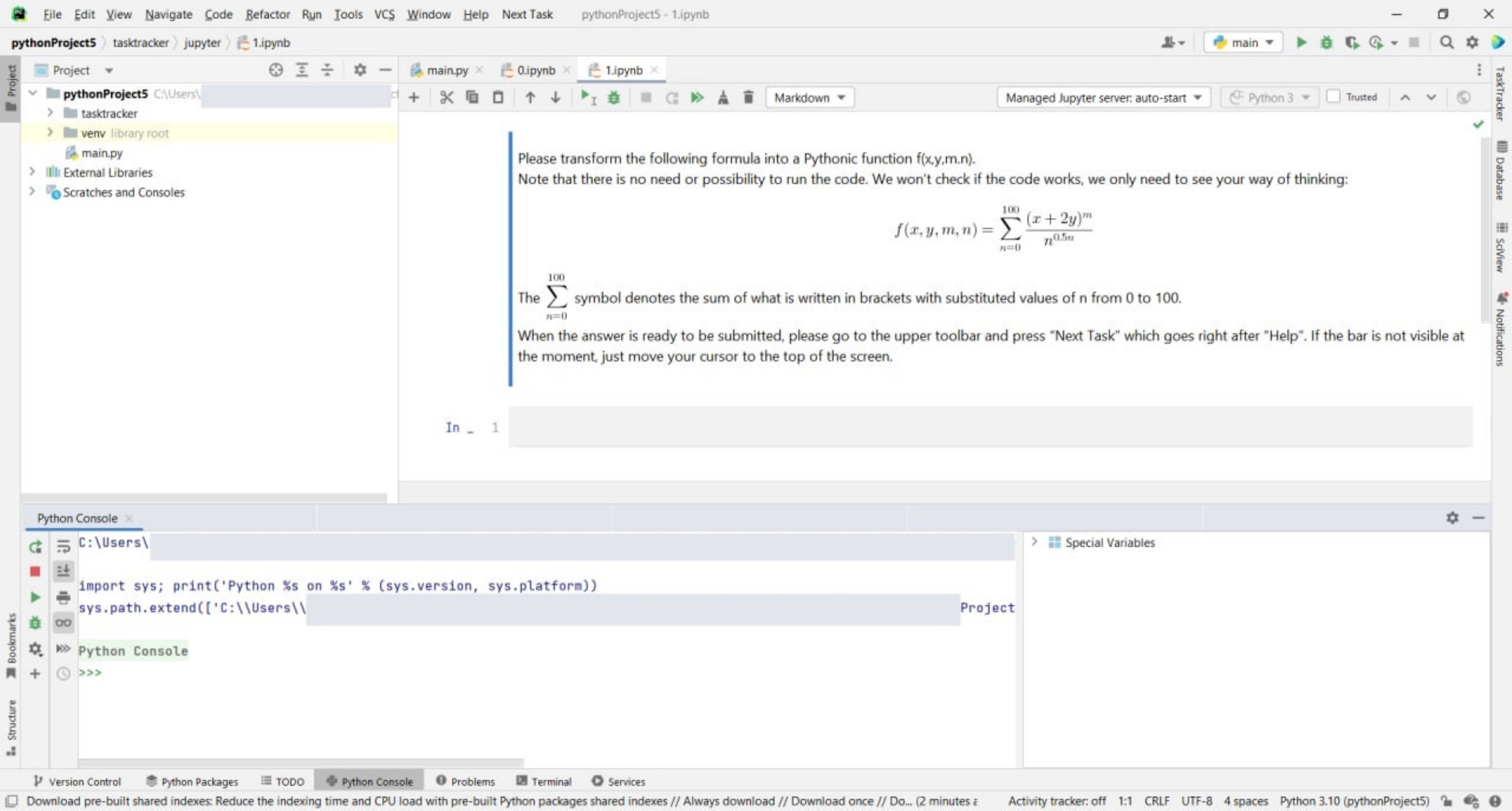}}
    \caption{The coding task in high perceptual load mode.}\label{fig:high}
    \subfloat{\includegraphics[width = 0.57\textwidth]{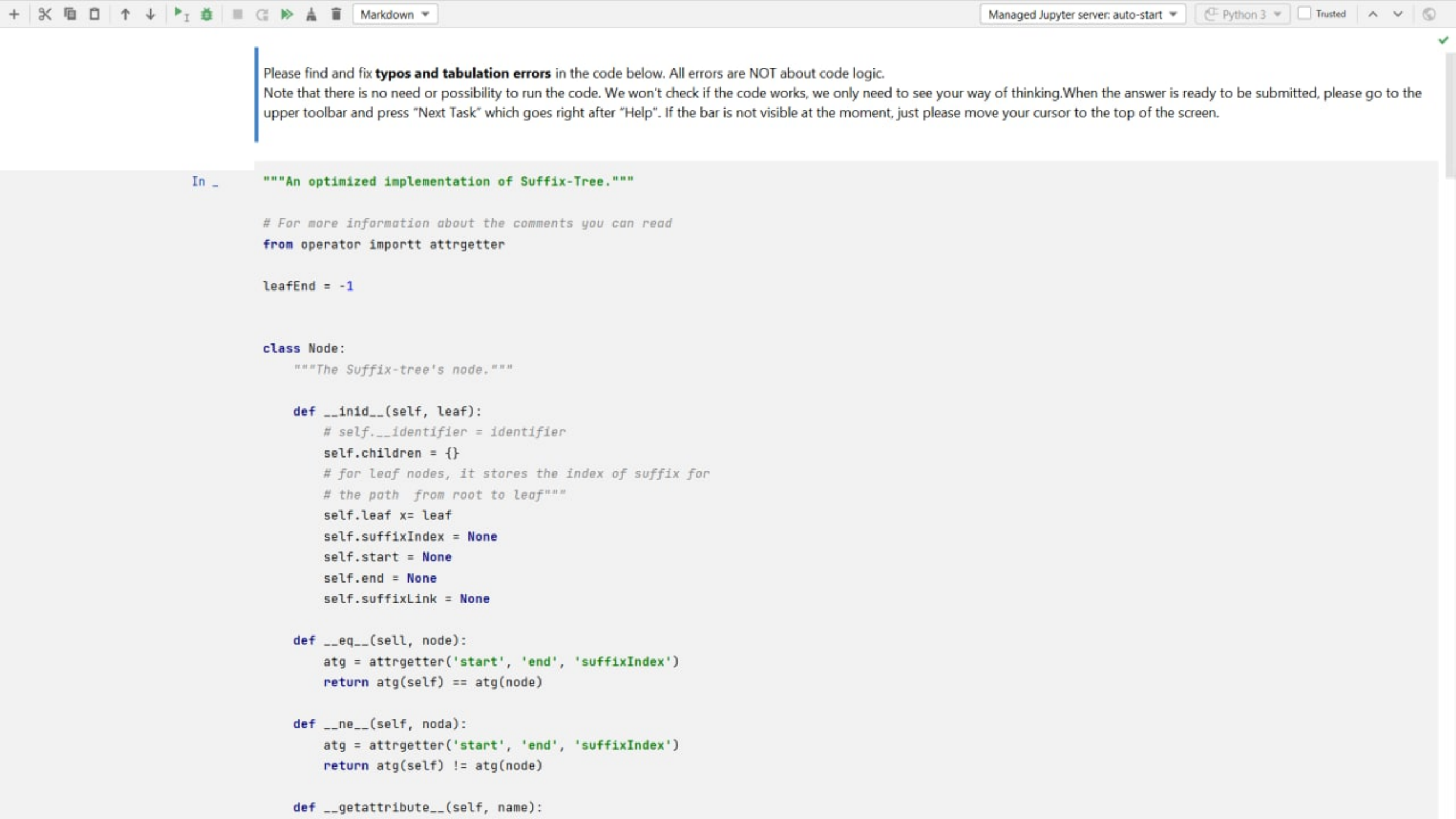}}
    \caption{The debugging task in low perceptual load mode.}\label{fig:low}
    \end{figure}

To emulate the programming activity, we used two types of tasks: (a) to write a programming function that performs the calculation in accordance with a given math formula and (b) to debug a code snippet that contains typos and syntactic errors. The tasks were chosen to be solvable by developers on any level of experience with the Python programming language. To test whether \PL affects the efficiency of the solution, we made two comparable tasks of every type to present them in \HPLM and \LPLM. Thus, we came up with 4 tasks of the same level of complexity --- two for coding and two for debugging (for full texts, see replication package~\cite{pack}).
 
In our research, the coding task was used to trigger a more mentally active state, while debugging was considered monotonous. To amplify those traits, we made the coding task relatively short and active --- the task triggered a problem-solving process and the solution was expected to be less than 10 lines long. 

Snippets for debugging on the other hand were long and focus-demanding --- they consisted of 186 and 196 lines of code with 32 typos to be fixed in each. This division of tasks on the basis of mental activeness allowed us to test if the experimental intervention has specificity for the efficiency of the solution of a monotonous programming task. 

\medskip
\textbf{IDE Plugin.}

The described tasks were presented with the help of a plugin for the IntelliJ Platform~\cite{kurbatova2021intellij}, which we created to manipulate the interface and collect data. To fit the means and goals of our research, we adapted the TaskTracker plugin~\cite{lyulina2021tasktracker}. 
This plugin was developed by E. Lyulina and colleagues to be used in IntelliJ-based IDEs. To display the user interface, it used JavaFX~\cite{jfx} technology, which is deprecated since IntelliJ Platform 2020.2. The original TaskTracker supported logging the progress of solving problems, but did not support changing the configuration of the IDE and was not compatible with the latest versions of the IDE. Therefore, the plugin required significant improvement to suit our research. 

Our version of the extension was developed in Kotlin. The overall implementation is a combination of a server and a plugin for the IntelliJ platform. The server part was built on Node.js. For the UI, instead of JavaFX, we used the Java Chromium Embedded Framework~\cite{JCEF}, which is a built-in browser with all features that are usual for browsers. 

The architecture of our approach consists of several components: \textbf{virtual server} --- sends tasks and configurations, receives and stores data; \textbf{server} --- transmits tasks and configurations to the controller and passes on logs to the virtual server; \textbf{controller} --- implements configurations, initiates data logging; \textbf{user interface} --- developing environment; \textbf{logger} --- tracks activity and all symbol-by-symbol changes in the solution; \textbf{IntelliJ Platform SDK} --- software development kit to support customization of the IDE needed for our experiment.

This architecture allowed us to perform the following tasks: (a) embed research-related UI into the IDE, (b) log the progress of problem solving, and (c) change the configuration (perceptual load) of the development environment. 

In our study, we operationalize perceptual load as the number of interactable panels in the IDE. As can be seen in Figure~\ref{fig:low}, the IDE in \LPLM has two visible panels (\textit{i.e.}, a logically and visually united block of buttons and clickable elements), while \HPLM has seven of them (see Figure~\ref{fig:high}). 

Research-related UI consisted of pages with the informed consent form, BDEFS-SF, and feedback form. Each user was randomly assigned to the group after completing the consent form. Depending on the group, the user's IDE configuration was initially set to low or high perceptual load mode. The group also affected the sequence of task types in each mode --- whether coding or debugging would be the first. After the first pair of coding and debugging tasks, the \PL switched to the opposite one where again the sequence of task's type was counterbalanced between the sample. That resulted in eight possible configurations of the tasks' sequence.

After the completion of the study, all data on the user's solutions, as well as the data received from the activity tracker was sent to the server.

To test real-world effects, we set up our experiment settings to be as close as possible to habitual ones --- using a production-grade IDE on a personal computer with real-world tasks helped us to do so. \LPLM is represented by a built-in ``Zen'' mode of the IntelliJ-based IDE PyCharm, this feature turns off the unneeded warnings and windows. \HPLM is IDE's default visual representation. Thus, we maintained ecological validity, \textit{i.e.}, the potential to make ``a generalization of the experimental findings to the real world outside the laboratory''~\cite{orne1968ecological}. The plugin only implemented several changes to the habitual IDE functioning to build a more formal experimental environment. Such functions as auto-completion, error inspections and highlighting of code duplicates were turned off to test the mental activity of the participants themselves and to maintain the monotony when needed.

\medskip

\textbf{Metrics.}

The plugin collected and recorded time, as well as all actions and every symbol change made by a user, \textit{i.e.}, indicators of solution efficiency. From this, we build two sets of metrics --- for coding and debugging, respectfully.

For the coding task, the measures were as follows:
\textbf{Total time taken to solve the task} (sec) --- time delta between the appearance of the task on the screen and the ``Next'' button press;
\textbf{Time of input of the first character} (sec) --- time delta between the appearance of the task on the screen and the first keystroke;
\textbf{Average time for one line of code} (sec) --- time delta between the first and last keystroke divided by the number of lines in the task's solution;
\textbf{Speed} (diff/sec) --- the number of changed characters in the solution (\textit{i.e.}, insertion of letters, deletions, newlines, tabs) divided by the total time spent on the solution;
\textbf{Number of deletions} --- the number of times when the symbol was deleted in the solution;
\textbf{Number of actions} --- the number of times special keys and shortcuts were used (\textit{i.e.}, Enter, Shift, Backspace, arrows, Ctrl+A, etc.);
\textbf{Number of lines in the solution};

 For the debugging task, the following indicators were tracked:
\textbf{Total time taken to solve the task} (sec) --- time delta between the appearance of the task on the screen and the ``Next'' button press;
\textbf{Time of the first bug fix} (sec) --- time delta between the appearance of the task on the screen and the first bug fix;
\textbf{Number of changes in the snippet} --- the number of changed characters in the initial snippet (\textit{i.e.}, insertion of letters, deletions, newlines, tabs) compared to the final snippet;
\textbf{Number of actions} --- the amount of usage of special keys and shortcuts;
\textbf{Number of fixed bugs}.

\subsection{Data Collection}

The experiment consisted of three steps, where participants:

1. installed the pre-configured plugin and completed the built-in BDEFS-SF questionnaire. There, they indicated how frequently they encounter difficulties in self-directed actions, \textit{i.e.}, executive functions.

2. were presented with two pairs of tasks in the IDE. Each pair consisted of one coding task and one debugging task. To counterbalance possible learning and fatigue effects, half of the sample was presented with the first pair of tasks in \LPLM, and the other half --- in \HPLM. After the second task, \PL of the environment changed to the opposite one, and a new pair of tasks was presented. In each mode, the sequence of appearance of coding and debugging tasks was counterbalanced between groups. 

3. received BDEFS-SF feedback and were asked to name the techniques (if any) they use to cope with self-regulation problems in their daily lives. Finally, they received questions about their months of experience with the Python language and familiarity with the ``Zen'' mode.

All experimental sessions were administered online via video conference with screen sharing from the participant's side. This helped us (a) to notice and control environmental variables such as level of noise, participant's posture, etc; (b) to make sure everybody in the sample understand the flow and tasks of the experiment comparatively at the same level; (c) to troubleshoot immediately if needed; (d) to ensure that participants completed the tasks bonafide.

\subsection{Data Analysis}

All data obtained was analyzed in Python. Statistical analysis was performed with Scipy~\cite{virtanen2020scipy}, NumPy~\cite{harris2020array}, Pandas~\cite{mckinney-proc-scipy-2010}, and Pingouin~\cite{vallat2018pingouin}. 
In our research, the dependent variables are the solution's efficiency measures. The independent variables were represented by: (a) impairment of executive functions (presence or absence of the symptom) and (b) IDE's perceptual load (high and low). 

The decision on the presence or absence of the symptom was made on the grounds of the overall factor's score --- if a participant had a score greater than 8 on the BDEFS-SF factor under investigation, they were considered as having that ADHD symptom.

The results of the survey did not appear to be significantly correlated with the efficiency measures of the tasks according to performed Spearman’s pairwise correlation with Bonferroni corrections. 

The statistical hypotheses of our research to answer the research questions for coding and debugging programming tasks are presented further.

\medskip
\textbf{Hypothesis 1.1 (RQ1) There is a difference in efficiency measures depending on the perceptual load of the IDE.}

To test whether \PL of the programming environment affects the efficiency of task solving, we used the paired-samples T test~\cite{zimmerman1997teacher} to assess the difference in efficiency metrics between tasks solved in high and low \PL for all participants without dividing them into groups.

\medskip
\textbf{Hypothesis 2.1 (RQ2) Perceptual load mode and impairment of executive functions together affect the efficiency of the task's solution.}

When the effect of \PL was confirmed, to investigate the effect of \PL across people with and without impaired executive functions, we performed a series of two-way repeated measures analysis of variance (ANOVA)~\cite{girden1992anova}. We tested whether the interaction of (a) IDE's \PL and (b) impairment of executive functions affects the efficiency metrics. 

\medskip
\textbf{Hypothesis 2.2 (RQ2) There is a difference in the ratio of efficiency for people with and without ADHD symptoms.}

After the interaction of factors was confirmed, we used the Mann–Whitney U test~\cite{mcknight2010mann} to see how exactly the \PL effect differs for people with and without ADHD symptoms. For that, we calculated \emph{the ratio of efficiency}, which is an indicator of the \PL effect normalized to the one's results.

\emph{The ratio of efficiency} is calculated as the proportion of the efficiency measures in \HPLM to those in \LPLM. Let us provide an example of its calculation. For instance, if the efficiency measure in question is the number of actions in debugging task, \emph{the ratio of efficiency} would be calculated for two groups --- with and without an ADHD symptom (\textit{e.g.}, with difficulties in Self-restraint). The ratio would be equal to the proportion of the number of actions in \HPLM to those in \LPLM. That might result in \(300 \div 400 = 0.75\) for those without the symptom and \(963 \div 321 = 3\) for those with the symptom. If the ratio is greater than 1, the measure was greater in \HPLM. Accordingly, if the ratio is less than 1, the measure was greater in \LPLM. This makes this measure an indicator of the \PL effect. The ratio of efficiency allows us to consistently compare the obtained data and to draw conclusions about the effects of \PL on the efficiency of programmers with different symptoms of ADHD.

For all statistical tests, we used an alpha level of .05. 

Testing the stated hypotheses provided insights into the usefulness of the change of IDE's perceptual load for neurodiverse people in software engineering for mentally active and monotonous programming tasks.

\begin{table}[ht]
\centering
\caption{Descriptive statistics for Coding efficiency measures}
\begin{adjustbox}{width=\textwidth}
\begin{threeparttable}

\begin{tabular}{lcccccccc}
\toprule

&   \textbf{PL*}   &     \textbf{mean} &     \textbf{std} &     \textbf{min} &     \textbf{25\%} &     \textbf{50\%} &     \textbf{75\%} &      \textbf{max} \\
\midrule
 \rowcolor{gray}
\multirow[c]{2}{*}{Total time (sec)} &  High &   226.39 & 107.08 & 87.14 & 158.06 & 201.39 & 264.05 & 583.89 \\
 &  Low &   208.74 & 128.09 & 89.08 & 121.44 & 170.31 & 247.13 & 634.65 \\
 \rowcolor{gray}
\multirow[c]{2}{*}{Number of actions} &  High &   68.56 & 40.45 & 26 & 43 & 59.5 & 75.25 & 189 \\
 &  Low &   77.44 & 60.22 & 2 & 41.25 & 66.5 & 86 & 326 \\
 \rowcolor{gray}
\multirow[c]{2}{*}{Time of input of the first character (sec)} &  High &   34.68 & 26.3 & 2.36 & 10.88 & 31.2 & 50.27 & 112.01 \\
 &  Low &   22.53 & 13.47 & 4.61 & 12.29 & 19.26 & 27.17 & 63.21 \\
 \rowcolor{gray}
\multirow[c]{2}{*}{Average time for one line of code (sec)} &  High &   54.77 & 28.86 & 16.55 & 32.6 & 46.19 & 78.22 & 115.95 \\
 &  Low &   50.38 & 42.49 & 12.9 & 23.85 & 33.06 & 62.55 & 206.07 \\
 \rowcolor{gray}
\multirow[c]{2}{*}{Number of deletions} &  High &   21.69 & 19.89 & 1 & 9 & 15 & 26.5 & 92 \\
 &  Low &   23.61 & 17.13 & 5 & 12.75 & 18 & 29.75 & 8 \\
 \rowcolor{gray}
\multirow[c]{2}{*}{Number of lines in the solution} &  High &   4.75 & 1.93 & 2 & 3 & 5 & 6 & 9 \\
 &  Low &   5.25 & 2.37 & 1 & 4 & 5 & 7 & 12 \\
 \rowcolor{gray}
\multirow[c]{2}{*}{Speed (diff/sec)} &  High &   0.77 & 0.35 & 0.27 & 0.5 & 0.72 & 0.95 & 1.77 \\
 &  Low &   0.99 & 0.49 & 0.24 & 0.67 & 0.93 & 1.28 & 2.65 \\
 \bottomrule
\end{tabular}

\begin{tablenotes}
\footnotesize
    \item $*$PL --- perceptual load    
\end{tablenotes}

\end{threeparttable}

\end{adjustbox}

\label{table:desccode}
\end{table}

\begin{table}[ht]
\centering
\caption{Descriptive statistics for Debugging efficiency measures}
\begin{adjustbox}{width=\textwidth}
\begin{threeparttable}
\begin{tabular}{lcccccccc}
\toprule

&   \textbf{PL*}   &     \textbf{mean} &     \textbf{std} &     \textbf{min} &     \textbf{25\%} &     \textbf{50\%} &     \textbf{75\%} &      \textbf{max} \\
\midrule
\rowcolor{gray}
\multirow[c]{2}{*}{Total time (sec)} &  High &   879.67 &  432.99 &    3.01 &  588.93 &  847.41 &  1049.03 &  1762.88 \\
 &  Low &   776.19 &  304.95 &  245.56 &  573.75 &  765.36 &   932.72 &  1497.81 \\
\rowcolor{gray}
\multirow[c]{2}{*}{Number of changes in the snippet} &  High &   43.89 &   16.81 &   19.00 &   35.50 &   43.00 &    54.00 &   102.00 \\
 &  Low &   43.11 &   17.34 &   22.00 &   30.50 &   40.00 &    47.50 &    98.00 \\
\rowcolor{gray}
\multirow[c]{2}{*}{Number of fixed bugs} &  High &   22.54 &    5.76 &    6.00 &   20.00 &   24.00 &    27.00 &    30.00 \\
 &  Low &   21.77 &    5.70 &    2.00 &   18.50 &   23.00 &    26.00 &    30.00 \\
\rowcolor{gray}
\multirow[c]{2}{*}{Number of actions} &  High &   370.37 &  611.06 &   26.00 &   62.50 &  131.00 &   321.00 &  3101.00 \\
 &  Low &    425.03 &  653.51 &   20.00 &   70.50 &  141.00 &   560.50 &  3085.00 \\
\rowcolor{gray}
\multirow[c]{2}{*}{Time of the first bug fix (sec)} &  High &   61.87 &   42.72 &   12.82 &   32.62 &   54.78 &    75.97 &   189.90 \\
 &  Low &   45.15 &   36.98 &   11.42 &   18.88 &   35.62 &    56.79 &   181.26 \\
 \bottomrule
\end{tabular}

\begin{tablenotes}
\footnotesize
    \item $*$PL --- perceptual load    
\end{tablenotes}

\end{threeparttable}

\end{adjustbox}

\label{table:descdeb}
\end{table}

\begin{table}[ht]
\centering
\caption{Descriptive statistics for the BDEFS-SF results}
\begin{adjustbox}{width=\textwidth}
\begin{threeparttable}
\begin{tabular}{lrrrrrrrrr}
\toprule
&   \textbf{symptom*}   &     \textbf{mean} &     \textbf{std} &     \textbf{min} &     \textbf{25\%} &     \textbf{50\%} &     \textbf{75\%} &      \textbf{max} \\
 \midrule
Time management  & 21 & 10.06  & 3.24  & 5 & 8 & 10.5 & 12 & 16 \\
Self-organization  & 14 & 7.67  & 2.15  & 4 & 6 & 7 & 1 & 12 \\
Self-restraint  & 8 & 7.44  & 2.79  & 4 & 5 & 7 & 8 & 14 \\
Self-motivation  & 3 & 11.17  & 3.04  & 7 & 9 & 1 & 12.25  & 19 \\
Self-regulation of emotions  & 4 & 6.42  & 2.29  & 3 & 5 & 6 & 8 & 12 \\
\bottomrule
\end{tabular}

\begin{tablenotes}
\footnotesize
    \item $*$Symptom --- number of people who have such a symptom    
\end{tablenotes}

\end{threeparttable}

\end{adjustbox}

\label{table:descscale}
\end{table}
\section{Results} \label{sec:Results}

We hypothesize that for active programming tasks (coding) and monotonous ones (debugging), the perceptual load of the IDE might affect the time spent on the task and the number of user actions. Also, this effect might be different across people with different levels of executive function impairment. In this section, we provide the results of testing these hypotheses on data from 36 Python developers of various levels of experience (min = 6 months, max = 10 years, Median $\approx$ 3.5 years), who mostly never use the \LPLM of the IDE, the Zen mode, while working (35 out of 36 people did not use this mode, 10 out of 36 have known about the option).

Descriptive statistics for coding and debugging efficiency measures, as well as for BDEFS-SF, are presented in~\Cref{table:desccode,table:descdeb,table:descscale}. 

\smallskip
\textbf{RQ1. Does \PL of the programming environment affect the efficiency of task solving?}

\textbf{Hypothesis 1.1} \textit{There is a difference in the efficiency measures depending on the perceptual load of the IDE. }

The difference in efficiency metrics between tasks solved in high and low \PL was assessed using the paired-samples T test~\cite{zimmerman1997teacher}. 

\begin{figure*}[ht]
    \centering
    \includegraphics[width=0.8\textwidth]{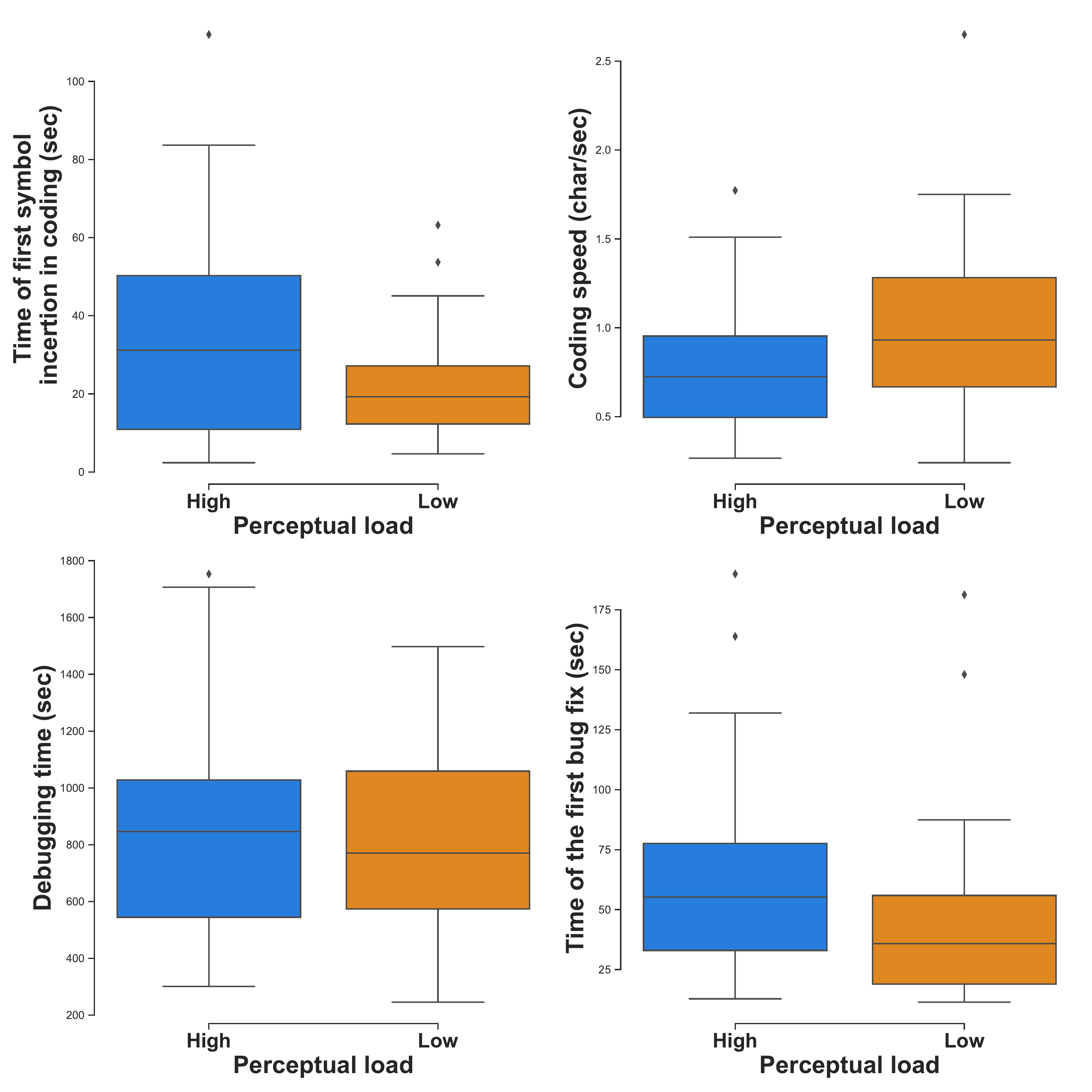}
    \caption{Box plots of the effect of the perceptual load on the efficiency measures.}
    \label{fig:ef_mode}
\end{figure*}

\smallskip
\textbf{For the coding task}:

We found a significant difference in the time of inserting the first character.  In \LPLM \APAstats{22.53}{13.47} it was less than in \HPLM \APAstats{34.68}{26.3}, \APAt{35}{-2.743}{.005}{.582}. 

We found that the coding speed in \LPLM \APAstats{0.99}{0.49} was significantly higher than in \HPLM \APAstats{0.77}{0.35}, \APAt{35}{2.717}{.005}{.532}). 

\smallskip
\textbf{For debugging tasks}:

The overall debugging time was significantly different between modes --- less in \LPLM \APAstats{780.34}{324.2} than in \HPLM \APAstats{885.29}{400.95}, \APAt{35}{-1.866}{.035}{.288}.

Time when the first bug was found significantly less in \LPLM \APAstats{45.41}{36.48} than in \HPLM \APAstats{62.96}{42.52}, \APAt{35}{ -1.778}{.042}{0.44}.

Other efficiency measures did not show any statistical significance in the difference between modes. 

\observation{In our research, the perceptual load did affect some of the solution efficiency measures (see Figure~\ref{fig:ef_mode}). \\For coding, people tend to start solving the task earlier, and the speed of writing the code was higher in \LPLM, compared to \HPLM. For debugging, the overall time taken and the time when the first bug was fixed were less in \LPLM, compared to \HPLM. \\When the effect of \PL was confirmed, we investigated to see if it differs between people with and without ADHD symptoms.}

\textbf{RQ2. Does the effect of \PL differ between people with and without impaired executive functions?}

\textbf{Hypothesis 2.1} \textit{Perceptual load mode and the impairment of executive functions together affect the efficiency of the tasks' solution.}

To test whether the factors of the IDE's \PL and ADHD symptoms together affect the efficiency measures of task solving, we performed a series of two-way repeated measures analysis of variance (ANOVA)~\cite{girden1992anova}.   

\smallskip
\textbf{In coding:}

The time of input of the first character was affected by the perceptual load and Self-restraint impairment together \APAf{1}{34}{5.869}{.021}{.147}. 

The number of lines in the solution appeared to depend on difficulties with Self-motivation combined with the perceptual load as factors \APAf{1}{34}{5.656}{.023}{.143}.

\smallskip
\textbf{In debugging:} 

The time when the first bug was fixed was affected by the interaction of \PL and the presence of Time management executive function impairment (\APAf{1}{34}{7.441}{.01}{.18}). 

The time of the first bugfix was also affected by the interaction of \PL with difficulties in Self-motivation (\APAf{1}{34}{5.726}{.022}{.144}).

\begin{figure*}[ht]
    \centering
    \includegraphics[width=0.65\textwidth]{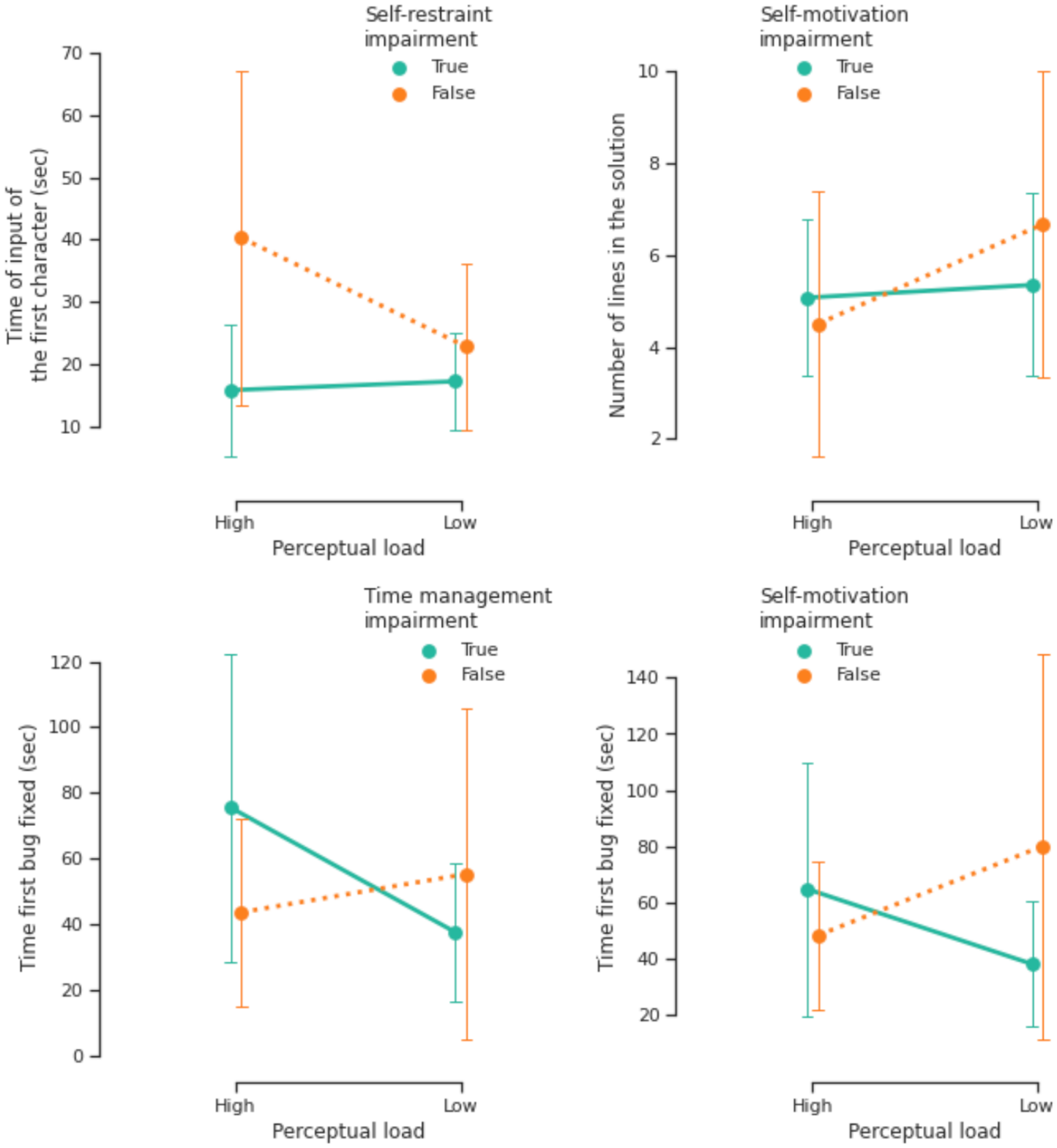}
    \caption{Plot of the dependence of task's solution efficiency from the interaction of ADHD symptom and the Perceptual load}
    \label{fig:anova}
\end{figure*}

Visualization of the effect of \PL and impairment of executive functions on efficiency metrics can be found in Figure~\ref{fig:anova}.

\observation{In our research, people with and without executive function impairments are affected differently by the perceptual load. \\We were able to register a statistically significant effect of the interaction of \PL and ADHD symptoms on efficiency measures. However, to see the specificity of that effect, further tests were needed.}

\textbf{Hypothesis 2.2} \textit{There is a difference in the ratio of efficiency for people with and without ADHD symptoms.}

To see how the  \PL effect, expressed in the ratio of efficiency, differs for people with and without ADHD symptoms,  we used the Mann–Whitney U test~\cite{mcknight2010mann}.  To represent \emph{the ratio of efficiency}, the proportions of efficiency measures on \HPLM to those on \LPLM were calculated. If the ratio is greater than 1, the measure was greater in \HPLM. Accordingly, if the ratio is less than 1, the measure was greater in \LPLM.

\smallskip
\textbf{In coding:} 

The efficiency ratio of time when the first symbol was entered was significantly less for the group of those who have difficulties with Self-restraint \APAstats{0.9}{0.62}, than for the group without that executive function impairment \APAstats{2.07}{1.64}, \APAu{57}{.02}{.746}. This could be interpreted as the fact that people with Self-restraint difficulties were almost equally fast in \LPLM and \HPLM, while people without such symptom were more active in \HPLM compared to \LPLM.  

\smallskip
\textbf{In debugging:} 

Only for the group with time-management difficulties, the proportion of the time when the first bug was found was greater \APAstats{2.62}{2.12} than for those with no ADHD symptoms \APAstats{1.41}{1.14}, \APAu{210}{.048}{.667}. For the people with time-management difficulties, it took longer to find and fix the first bug in \HPLM than in \LPLM. The same is true for people without such impairment. On the other hand, for people with symptoms, that efficiency proportion was significantly greater.

For the groups with other impairments of executive function, the efficiency ratio of the measurements was mostly less than 1 (see Table~\ref{table:2}), which means that the efficiency measure was greater in \LPLM than in \HPLM. In contrast, for the groups without corresponding impairments, the ratio was greater than 1. This might be interpreted as that in our research, people with ADHD symptoms were more active on the \LPLM when solving the monotonous task, compared with \HPLM, and at the same time, people without those symptoms were more active in the \HPLM. 

\begin{table}[ht]
\centering
\caption{Results of the Mann-Whitney U test for Debugging task}
\begin{adjustbox}{width=\textwidth}
\begin{threeparttable}
\begin{tabular}{llccccc}
    \toprule
    \multirow{2}{*}{\textbf{ADHD symptom}} & 
    \multirow{2}{*}{\textbf{Efficiency Measure}} & 
    \multicolumn{2}{c}{\textbf{Efficiency ratio for group}} 
    & \multirow{2}{*}{\textbf{U}} 
    & \multirow{2}{*}{\textbf{p}}
    & \multirow{2}{*}{\textbf{CLES}} \\
    &  & \multicolumn{1}{c}{\textbf{with S*}} & {\textbf{without S*}} & & & \\ 
    \midrule
    
    \rowcolor{gray}
    {Time management}
    & Time of first bug fix (sec) & \multicolumn{1}{c}{2.62 ± 2.11} & 1.41 ± 1.14 & 210 & .048 & .667\\
    
    {Self-restraint}     
    & Number of actions & \multicolumn{1}{c}{0.61 ± 0.44} & 1.31 ± 0.93 & 59 & .022 & .737\\
    
    \rowcolor{gray} 
    {Self-organization} 
    & Number of changes in the snippet & \multicolumn{1}{c}{0.97 ± 0.35} & 1.2 ± 0.45 & 98 & .036 & .682\\ 
    
    \multirow{2}{*}{Self-regulation of emotions} 
    & Total time (sec) & \multicolumn{1}{c}{0.84 ± 0.15} & 1.25 ± 0.44 & 26 & .03 & .797\\ 
    & Number of changes in the snippet & \multicolumn{1}{c}{0.7 ± 0.2} & 1.16 ± 0.42 & 19 & .01 & .852 \\
    \bottomrule

\end{tabular}

\begin{tablenotes}
\footnotesize
    \item *S --- the symptom
    \item U --- Mann-Whitney test statistic, reflecting the difference between groups
    \item p --- probability value, describing how likely it is that the results would have occurred by random chance
    \item CLES --- Common Language Effect Size, the effect size of the difference between groups
    
\end{tablenotes}
\end{threeparttable}
\end{adjustbox}
\label{table:2}
\end{table}

\observation{Our results support the idea that people with and without executive function impairments are affected differently by the perceptual load. \\For coding and debugging, the mean efficiency ratio was mostly not greater than 1 for those who have symptoms of ADHD, which means that the efficiency measure was greater in the \LPLM.}  
\section{Threats to Validity} \label{sec:ttv}

In our research, we found statistically significant changes in the measures of the efficiency of solving tasks, related to the perceptual load of the IDE. This change was different for mentally active programming tasks (coding) and monotonous ones (debugging), and it also differs across people with different levels of impairment of executive functions.

Several limitations of the study are important to note. In our research, we study people with self-reported symptoms of ADHD. To determine the presence and the severity level of ADHD symptoms, we used a questionnaire, which is not enough to make any medical conclusions. This might cast a question of the generalizability of our results to those who have the diagnosis and specific neurodevelopmental issues. In future work, it is important to work with ADHD-diagnosed people. However, finding a large enough sample of diagnosed people who are also proficient in coding could be a challenging task. 

In the sample we managed to obtain, there is a variation in symptom-group sizes, some of them are relatively small. While that has no effect on the statistical analysis we perform due to the absence of relevant assumptions for the tests, the larger sample would possibly make effect sizes more notable.

Another limitation of the generalization of results is due to the changes we made to the IDE functionality for this study. To make the tasks independent of the participants' experience of using IDE and to make them challenging, we turned off some features that people use in IDE on a regular basis: autocompletion, error inspections, and highlighting of code duplicates. Enabling such features would make tasks as those presented in the study too easy. However, our goal was to test the effect of the perceptual load not on simple or difficult tasks, but on mentally active and passive ones. Thus, disabling hints helped us do this on smaller tasks. For this reason, we believe that the results of the study can still generalize to more complex tasks with the enabled features of the IDE if they are still mentally active or monotonous for developers.

The same argumentation is also applied to the design of the tasks --- suggested tasks significantly differ from the ones that professional developers meet in their day-to-day practice. However, we chose to present simpler cases of programming and debugging for the sake of independence from the participants' skill levels. As well as for the IDE changes, we believe that tasks used in our research could accurately represent more complex, real-life cases. These limitations could be solved by balancing the experimental sample by skill level and experience working in the IDE. However, finding such a sample is not trivial and is left for future work.
\section{Discussion} \label{sec:discussion}

Acknowledging the limitations listed above, we conclude that the perceptual load does affect the solution efficiency of programming tasks and that this effect differs across people with and without ADHD symptoms. 

In our research, \LPLM proved to be beneficial in terms of the speed of task completion in both mentally active and monotonous tasks. 

We found that programmers' efficiency might be affected by the interaction of the \PL and executive functions' impairment. This means that the effect of these factors together differs between people with various levels of severity and types of ADHD symptoms.

Our data shows that, while solving a monotonous programming task: (a) people with difficulties in Self-organization, Self-restraint, and Self-regulation of emotions might be more active in the \LPLM environment; (b) people who struggle with Self-regulation of emotions might get time-wise benefits from \HPLM;  (c) people with difficulties in time management benefit from \LPLM if the measure of efficiency is speed.

This makes our findings somewhat controversial. However, the important note here is the division of ADHD symptoms into clusters of impairments of executive functions in our work. The results of our experiments might be interpreted as confirming that people with some symptoms or with the overall severity of ADHD will indeed benefit from the high perceptual load as shown in previous work. At the same time, in some cases, where the severity of ADHD symptoms is dictated by a specific executive function impairment, for instance, time management, as in our research, people may still benefit from the low perceptual mode for monotonous tasks. This supports Barkley's idea of domains of ADHD symptoms, which might cause their specificity and therefore emphasize the need for different accommodations. 
\section{Conclusion}
There is a trend in the IT field to hire more neurodiverse people due to their special advantages and unique traits, which are beneficial in terms of innovation and profit. This presents a problem of accommodating such a workforce and helping them to cope with their day-to-day work life. One possible workplace adjustment is changing the work-station, which in software engineering might be presented by an IDE. Considering related work, we conducted research on how the perceptual load of the software interface might support developers with ADHD symptoms being efficient in their job. In our experimental research, 36 developers filled out the scale of ADHD symptoms and solved mentally active and monotonous programming tasks on high and low perceptual load modes of the IDE. Overall, our findings support the idea of perceptual load effect on the programmers' productivity. The results also show that people with ADHD symptoms might benefit from the low or high perceptual load of the environment, depending on the core specificity of the symptom. Therefore, we think that it would be useful to develop the IDE with built-in appearance adjustment tools to support the professional performance of neurodiverse people.

\bibliographystyle{splncs04}
\balance
\bibliography{IEEEabrv,paper}

\end{document}